\documentclass[lettersize,journal]{IEEEtran}
\usepackage{amsmath,amsfonts}
\usepackage{algorithmic}
\usepackage{algorithm}
\usepackage{array}
\usepackage[caption=false,font=normalsize,labelfont=sf,textfont=sf]{subfig}
\usepackage{textcomp}
\usepackage{stfloats}
\usepackage{url}
\usepackage{verbatim}
\usepackage[none]{hyphenat} % 关闭自动断字
\usepackage{graphicx}
\usepackage{cite}
\begin{document}

\title{Sparse Arrays Enable Near-Field Constant-Distance Focusing with Reduced Focal Shift}

\author{Jiawang Li, Yingjie Xu,~\IEEEmembership{Student member, IEEE}, Hanieh Aliakbari,~\IEEEmembership{Senior Member, IEEE}%
\thanks{Manuscript received xxxx.xx.xx. (Corresponding author: \textit{Jiawang Li}).}%
\thanks{Jiawang Li, Yingjie Xu, Hanieh Aliakbari are with the Department of Electrical and Information Technology, Lund University, 22100 Lund, Sweden (e-mail: jiawang.li@eit.lth.se).  Hanieh Aliakbari is also with Volvo Car Corporation, 
Gothenburg, Sweden. }%
\thanks{Color versions of one or more of the figures in this letter are available online at http://ieeexplore.ieee.org.}%
\thanks{Digital Object Identifier 10.1109/AWPL.2025.xxx.}
}

% The paper headers
\markboth{Journal of \LaTeX\ Class Files,~Vol.~14, No.~8, February~2025}%
{Li: Sparse arrays enable near-field constant-height focusing with reduced focal shift and improved beam uniformity.}

\maketitle

\begin{abstract}
In near-field beam focusing for finite-sized arrays, focal shift is a non-negligible issue. The actual focal point often appears closer to the array than the predefined focal distance, significantly degrading the focusing performance of finite aperture arrays. Moreover, when the focus point is scanned across different locations, the degradation becomes even more pronounced, leading not only to positional deviation but also to substantial energy loss.
To address this issue, we revisit the problem from the perspective of communication degrees of freedom. We demonstrate that a properly designed sparse array with optimized element spacing can effectively mitigate focal shift while enabling stable control of the focusing height during beam scanning.
Simulation results based on dipole antennas with different polarizations and patch antennas validate our findings. Notably, with optimized inter-element distances, the energy distribution across focal points becomes nearly uniform, and highly accurate focusing positions are achieved.
\end{abstract}


\section{Introduction}
\IEEEPARstart{L}{arge} intelligent surface (LIS) technology [1] has emerged as a promising solution to meet the stringent requirements of 6G networks in the sub-10 GHz bands, where spectrum resources are limited. As illustrated in Fig.1, numerous LIS modules are deployed along the interior walls, with inter-module communication capabilities. Each LIS module is composed of a large number of antenna elements. These elements collaboratively generate beams to enable high-capacity indoor wireless communications or efficient wireless power transfer.

In this scenario, the electromagnetic waves generated by the LIS through beamforming are no longer far-field plane waves [2], but should instead be treated as near-field spherical waves [3]-[5]. Consequently, conventional beamforming techniques [6]-[7] designed for far-field conditions are no longer applicable, which has led to increasing research attention on beamforming approaches specifically tailored for the near-field region [8]-[13].

These beamforming methods can be broadly categorized into two types. The first type is based on the conjugate-phase method [8]-[12], while the other is based on various optimization algorithms [13]. However, optimization-based methods often suffer from high computational complexity, a tendency to converge to local optima, and even convergence failure. As a result, conjugate matching methods are relatively more favored. In the design of finite-size arrays, particularly in practical LIS deployments, each antenna element is connected to an independent amplitude and phase control channel, making it difficult to scale up the size of individual LIS modules. Under such circumstances, the system inevitably encounters the focal shift problem as discussed in [8], especially when the LIS modules are deployed in strip-like configurations.

The near-field focal shift problem indicates that for finite-sized arrays, when employing conjugate matching methods to generate a near-field focus, the actual power peak often deviates from the intended focal point and tends to shift closer toward the array plane. This phenomenon can be explained as follows: in near-field focusing, the radiated power peak shifts toward the antenna because the amplitude increases with proximity, while phase coherence deteriorates. The stronger nearby amplitudes overcompensate for the phase mismatches, causing the maximum energy to appear before reaching the designated focal position.
\begin{figure}[t!]
    \centering
    \includegraphics[width=1\linewidth]{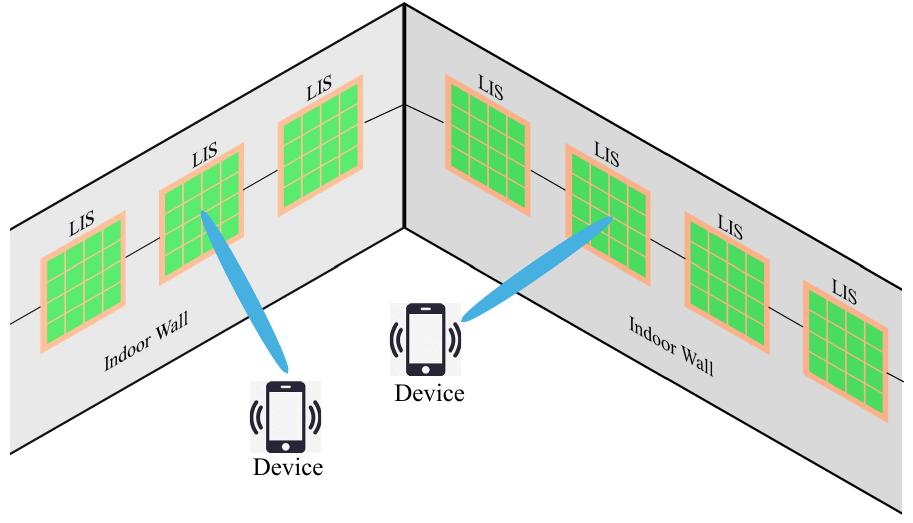}
    \caption{LIS application scenarios.}
    \label{fig:LIS_scenario}
\end{figure}

In far-field beam scanning, it is well known that the beam gain decreases as the scanning angle increases. Similarly, during near-field focusing with conventional arrays, severe focal shift and significant degradation of beam resolution occur. These issues can cause large positioning errors and may even prevent reliable communication with devices. Therefore, addressing focal shift and performance degradation in near-field beam scanning with finite arrays is crucial.

This paper analyzes the near-field focusing problem from the perspective of communication degrees of freedom and focuses on finite-sized one-dimensional linear arrays. It is shown that by properly setting the element spacing, the focal shift can be significantly reduced while maintaining a consistent peak power across different near-field focal points during scanning. To validate the proposed approach, electromagnetic simulations and verifications were conducted using dipole and patch linear arrays with different polarizations. The results demonstrate the effectiveness and practicality of the proposed method.

\begin{figure}[t!]
    \centering
    \includegraphics[width=1\linewidth]{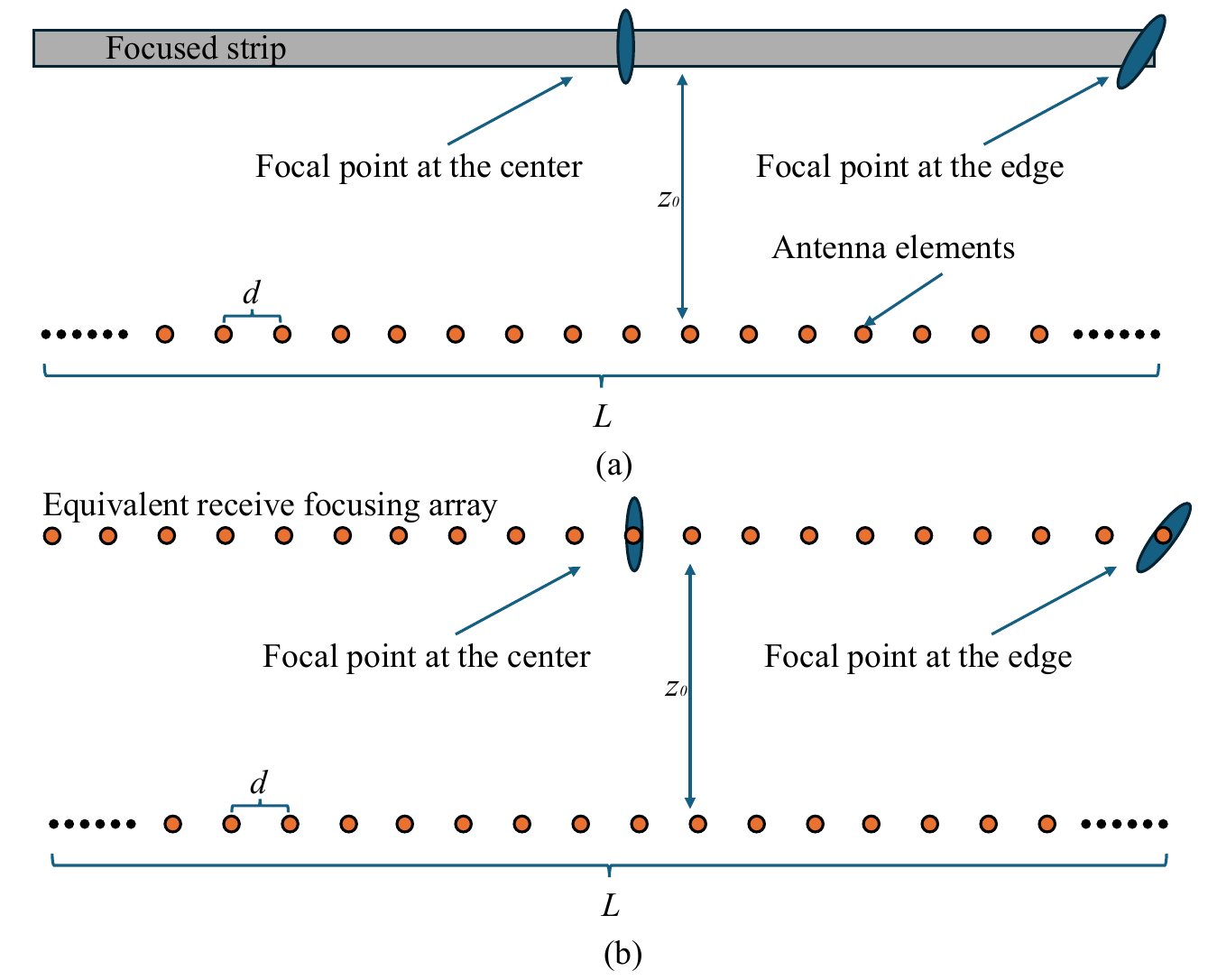} % Replace 'figure.png' with your image file
    \caption{Array configureation. (a) Linear array and the resulting focused strip distribution. (b) Equivalent parallel linear arrays. }
    \label{fig:example}
\end{figure}

\section{Array Configuration and Analysis}
As shown in Fig.~2(a), for a finite-sized linear array with an aperture length of $L$, the element spacing is set to $d$, where $d$ is not necessarily half a wavelength. The focused strip is located at a distance $z_0$ from the array, and its center is aligned with the center of the array. For simplicity, each antenna element is modeled as an isotropic radiator with its polarization oriented into the page. A more detailed analysis of polarization effects and array gain characteristics will be presented later. Our objective is to allocate the amplitudes and phases of the array elements such that, when the designated focal point is set at the strip center, the actual focus precisely coincides with this center position. Moreover, when the focal point is shifted toward the edges of the strip, the system should maintain stable positioning and consistent focusing power.

\subsection{Focusing Mechanism}
 First, it is important to understand that ensuring both the focal point at the center of the strip and the focal points near the strip edges are aligned around the central region essentially requires establishing two independent, high-quality communication channels. Consequently, the scenario depicted in Fig. 2 can be interpreted as communication between two parallel arrays. The problem of establishing as many independent channels as possible naturally leads to the concept of communication degrees of freedom (DoF) [14]-[16]. However, the focal points near the edges are more susceptible to misalignment and power loss due to the weaker contributions from distant antenna elements. As illustrated in Fig.~2(b), the receiving strip can thus be equivalently modeled as a receive array having the same aperture size as the transmit array (for simplicity, with identical element settings). Therefore, the original focusing problem can be reformulated as a maximization of the achievable DoF under this equivalent model.
The electric field on the receiving plane can be regarded as the superposition of contributions from all source points. Thus, based on the scalar Green's function, the electric field at the receiving point \( m \) can be expressed as [14]:
\begin{equation}
E(\mathbf{r}_{rm}) = \frac{1}{4\pi} \sum_{n=1}^{N} A_n \frac{\exp\left( -j k |\mathbf{r}_{rm} - \mathbf{r}_{sn}| \right)}{|\mathbf{r}_{rm} - \mathbf{r}_{sn}|}
= \sum_{n=1}^{N} A_n h_{mn}
\end{equation}
where \( k_0 \) is the free-space wavenumber, defined as \( k_0 = \frac{2\pi}{\lambda} \), with \( \lambda \) being the wavelength, and 
\begin{equation}
h_{mn} = \frac{1}{4\pi} \frac{\exp\left( -j k |\mathbf{r}_{rm} - \mathbf{r}_{sn}| \right)}{|\mathbf{r}_{rm} - \mathbf{r}_{sn}|}
\end{equation}
Here, \( \mathbf{r}_{rm} \) denotes the position vector of the \( m \)-th receiving antenna, 
and \( \mathbf{r}_{sn} \) denotes the position vector of the \( n \)-th transmitting antenna.
For the received signal \( \mathbf{f} = [f_1, f_2, \ldots, f_M]^T \) on the receiving strip, it can be obtained through the complex excitation vector \( \mathbf{A} = [A_1, A_2, \ldots, A_N]^T \) and the channel matrix \( \mathbf{H} \), where
\begin{equation}
\mathbf{H} =
\begin{bmatrix}
h_{11} & h_{12} & \cdots & h_{1N} \\
h_{21} & h_{22} & \cdots & h_{2N} \\
\vdots & \vdots & \ddots & \vdots \\
h_{M1} & h_{M2} & \cdots & h_{MN}
\end{bmatrix}
\end{equation}
In the current single-polarization scenario, the channel matrix is constructed by accumulating energy through a two-point method. The largest $N_e$ singular values dominate the matrix, while the remaining smaller singular values contribute negligibly to the overall degrees of freedom. Therefore, the degrees of freedom can be calculated as [14]:
\begin{equation}
n_{\text{e}} = \frac{\left( \mathrm{tr}\left( \mathbf{H}\mathbf{H}^H \right) \right)^2}{\left\| \mathbf{H}\mathbf{H}^H \right\|_F^2} = \frac{\left( \sum_i \sigma_i \right)^2}{\sum_i \sigma_i^2}.
\end{equation}
where ${\sigma_i}$ the $i$-th largest eigenvalue of the channel correlation matrix $\mathbf{H}\mathbf{H}^H$. Next, we investigate the conditions under which the degrees of freedom can be maximized, meaning that the transmitted waves are able to cover the receiving strip as fully as possible even during the focal point scanning process. One intuitive consideration is that when the focal peak generated by the transmit array is directed toward a specific antenna element, the sidelobe levels toward other elements, especially the two adjacent ones, should be minimized. Ideally, it should be possible to generate $M$ orthogonal basis functions at baseband that are individually focused onto $M$ different receive elements, thereby enabling independent receive channels. Therefore, it is meaningful to investigate how the focusing beamwidth varies with the number of antenna elements. By setting the center of the transmit linear array as the origin, the position of each point source can be expressed as:
\begin{equation}
x_n = \left( n - \frac{N + 1}{2} \right) d
\end{equation}
\begin{figure}[t!]
    \centering
    \includegraphics[width=0.75\linewidth]{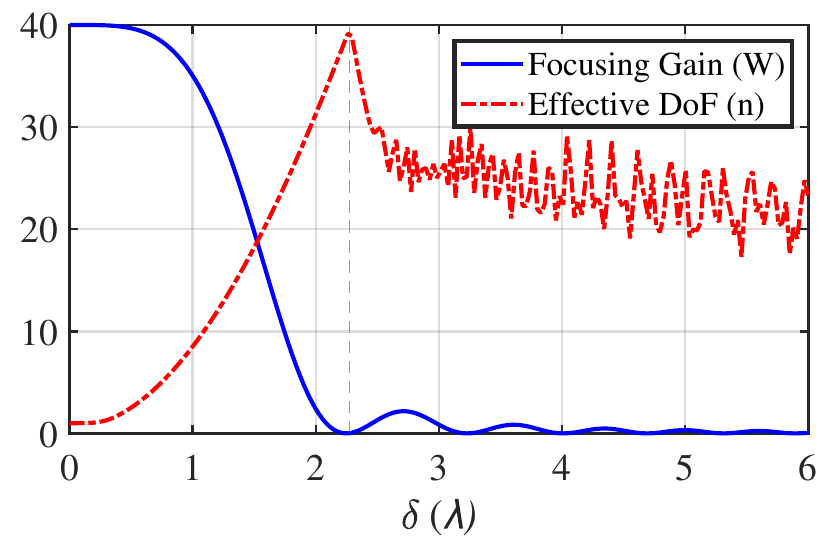} % Replace 'figure.png' with your image file
    \caption{Focusing gain and effetive DoF under the different $\delta$.}
    \label{fig:example}
\end{figure}
As shown in (1), considering practical scenarios, it is preferable to avoid adjusting the excitation amplitudes, since doing so would reduce the system's output efficiency. Instead, only the phases are adjusted to achieve conjugate matching and maximize power transfer. Therefore, the gain at the focusing position can be expressed as:
\begin{equation}
G = \frac{1}{N} \left| \sum_{n=1}^{N} \frac{e^{-j k \sqrt{(x_n - \delta)^2 + z_0^2}}}{4 \pi \sqrt{(x_n - \delta)^2 + z_0^2}} e^{j k \sqrt{x_n^2 + z_0^2}} \right|^2
\end{equation}
Here, $\delta$ denotes the offset from the focal peak. When the focal distance is significantly larger than the element position $x_n$, the amplitude variation caused by the denominator can be considered nearly constant, while the phase term remains unchanged [17]. Thus, the following approximation holds:
\begin{equation}
\frac{e^{-jk\sqrt{(x_n-\delta)^2 + z_0^2}}}{4 \pi \sqrt{(x_n-\delta)^2 + z_0^2}}
\approx
\frac{e^{-jk\sqrt{(x_n-\delta)^2 + z_0^2}}}{4 \pi z_0}
\end{equation}
Based on this condition, applying a Taylor expansion to equation (6) yields the following result:
\begin{equation}
G \approx \frac{1}{N} \left| \sum_{n=1}^{N} \frac{e^{-jk\left( z_0 + \frac{(x_n-\delta)^2}{z_0} \right)} e^{jk\left( z_0 + \frac{x_n^2}{z_0} \right)}}{4\pi z_0} \right|^2
\end{equation}
Substituting equation (5) into equation (8) and simplifying yields the following result:
\begin{equation}
G \approx \frac{1}{N} \left| \sum_{n=1}^{N} e^{j k \left( \frac{\delta}{2} \left( n - \frac{N+1}{2} \right) \delta \right) \frac{\delta}{z_0}} \right|^2
= N \frac{\operatorname{sinc}^2\left( \frac{N \delta^2}{\lambda L} \right)}{\operatorname{sinc}^2\left( \frac{\delta^2}{\lambda L} \right)}
\end{equation}
Therefore, when the spacing satisfies the following condition, a gain null is exactly achieved:
\begin{equation}
\delta = \pm \sqrt{\frac{n \lambda L}{N}}, \quad n = 1, 2, 3, \ldots
\end{equation}
In general, selecting the minimum spacing is preferable, as it helps avoid gain loss and grating lobe effects associated with excessively large spacings. As shown in Fig.~3, the focusing gain calculated using equation (9) and the effective degrees of freedom (DoF) obtained using equation (4) are presented. The simulation conditions are as follows: the operating frequency is 6~GHz, the focal distance $z_0 = 200\lambda$, and the array consists of 40 antenna elements. Therefore, under ideal conditions, the achievable DoF should be 40. When the maximum DoF is achieved, the element spacing is such that one element is located at the gain peak while the adjacent element is positioned at the first null. In the figure, the first null is located at approximately $2.27\lambda$.
Accordingly, equation (10) can be used to approximately compute the optimal spacing required for the focused array to fully cover the receiving strip. Naturally, if the focal distance $z_0$ is relatively small, the optimal spacing can be directly determined using equation (4).
\begin{figure}[t!]
    \centering
    \includegraphics[width=0.85\linewidth]{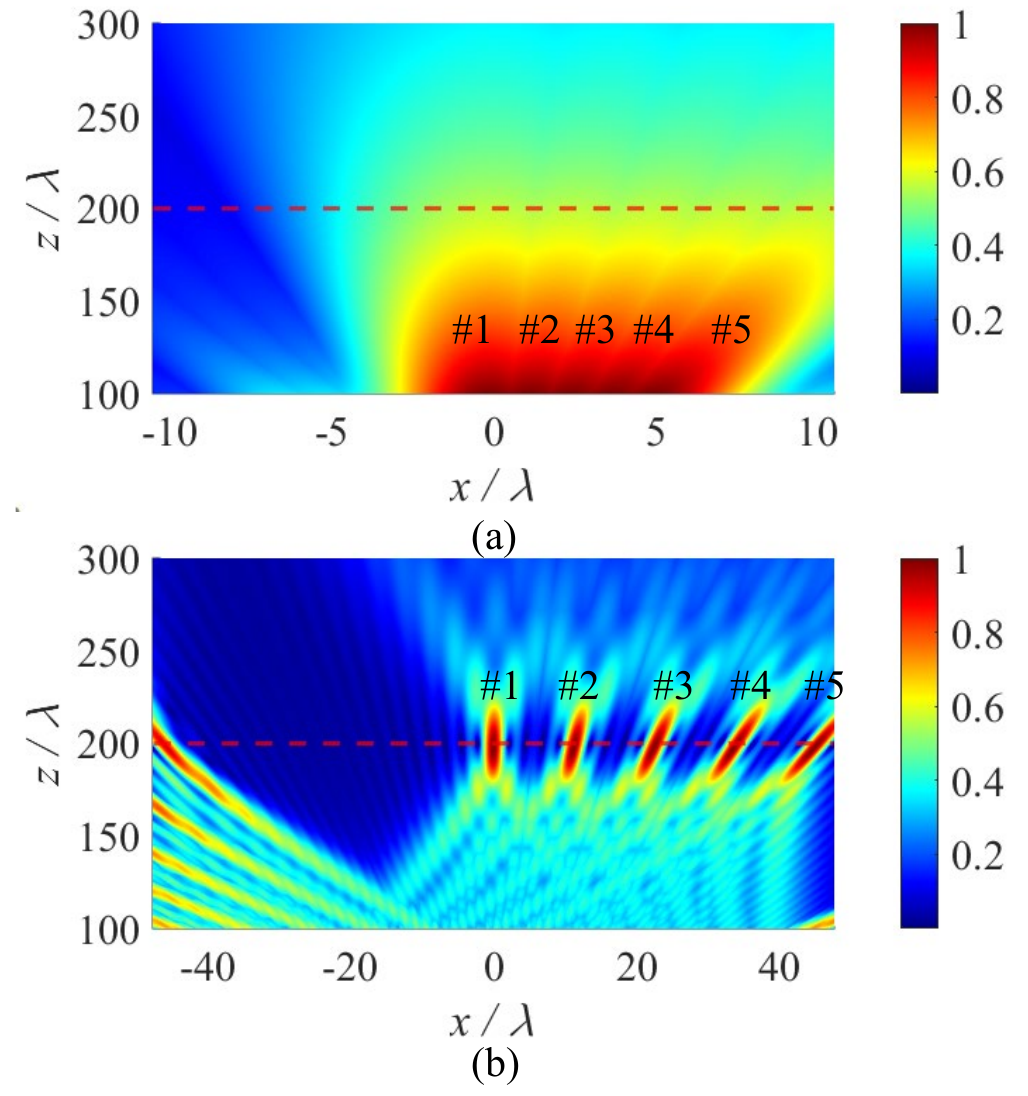} % Replace 'figure.png' with your image file
    \caption{Normalized focused electric field distributions for five scanning focal points with a spacing of $5\delta$: (a) $\delta = 0.5\lambda$ (conventional method). (b) $\delta = 2.27\lambda$
.}\label{fig:example}
\end{figure}
Fig.~4 presents the scanning results of the focused wave numbers generated by the proposed method (using the same parameters as in Fig.~3), along with a comparison to the conventional half-wavelength spacing approach. It is observed that, when a spacing of approximately $2.27\lambda$ is adopted, the focused position and energy remain stable even when scanning toward the edges of the array. In contrast, with the conventional half-wavelength spacing, the focal position deviates significantly from the intended target, making such an arrangement unsuitable for practical deployment. It is worth noting that when scanning toward the edge, specifically at the fifth focal beam, a relatively large sidelobe appears near the opposite edge of the receiving strip. However, the higher-order grating lobes associated with the other focal points do not fall within the focused strip region.

Next, we analyze the polarization characteristics of the proposed method. As shown in Fig.~5, we analyze three typical linear array configurations, including vertically polarized dipole elements, horizontally polarized dipole elements, and a patch antenna array with aperture distribution.
For a linear array composed of horizontally polarized dipole elements, the expression for the focused electric field is given by [18]:
\begin{figure}[t!]
    \centering
    \includegraphics[width=0.85\linewidth]{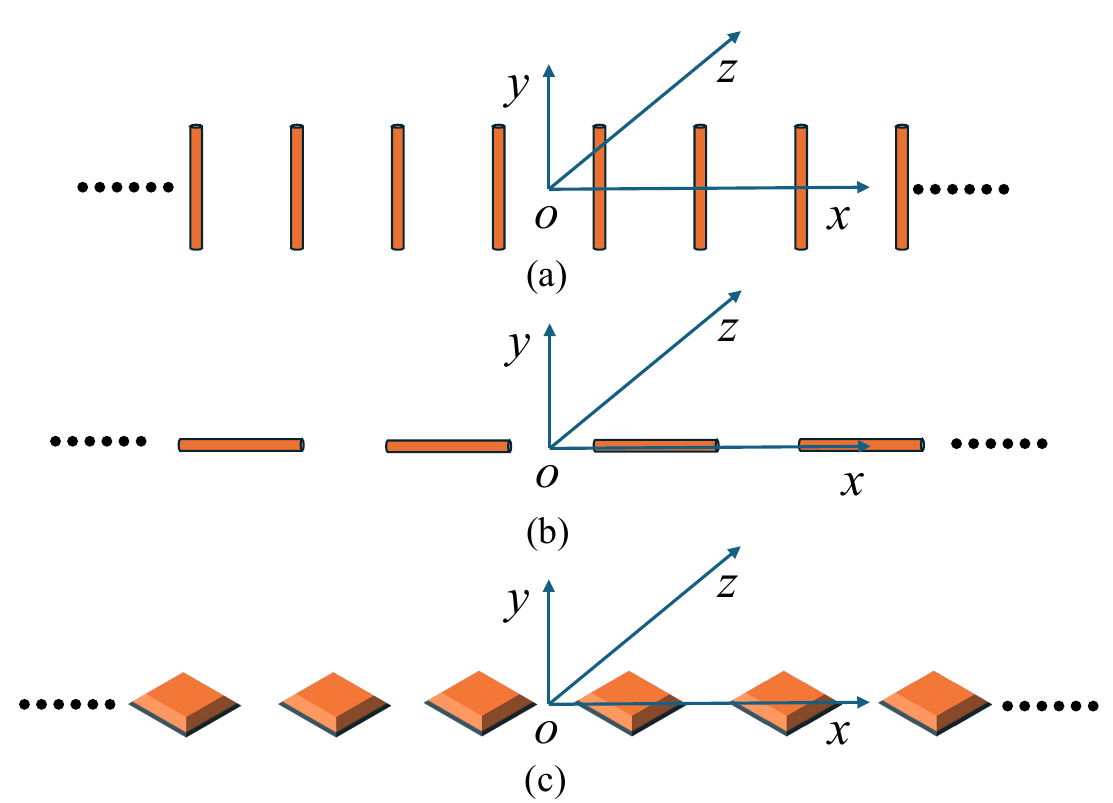} % Replace 'figure.png' with your image file
    \caption{Three types of array configurations: (a) vertically polarized dipole linear array. (b) horizontally polarized dipole linear array. (c) planar patch antenna array
.}\label{fig:example}
\end{figure}
\begin{equation}
E_x(z_0) = \sum_{n=0}^{N} \frac{z_0^2 e^{-jk\sqrt{z_0^2 + (x_n-\delta)^2}}}{\left( z_0^2 + (x_n-\delta)^2 \right)^{3/2}}
\approx
\sum_{n=0}^{N} \frac{e^{-jk\sqrt{(x_n-\delta)^2 + z_0^2}}}{z_0}
\end{equation}
Therefore, when focusing at relatively large distances, vertically and horizontally polarized dipole arrays exhibit similar characteristics.Next, we consider the planar patch array. A standard patch antenna exhibits a directional pattern proportional to $\cos(\theta)$. Taking the aperture effect into account, when the observation point deviates from the broadside direction of each patch, it is necessary to extract the projected power contribution from each patch. Therefore, based on equation (1), the expression must be further multiplied by the directional pattern function and the corresponding power projection, as follows:
\begin{equation}
E_p = E(\mathbf{r}_{rm}) \cos^2 \theta = E(\mathbf{r}_{rm}) \frac{z_0^2}{z_0^2 + (x_n-\delta)^2}
\end{equation}
The results yield an approximation consistent with equation (11), indicating that at relatively large focal distances, the three typical antenna types exhibit similar focusing characteristics. Naturally, as the focal distance becomes smaller, the differences between the two polarization configurations become increasingly evident.

\subsection{FEKO Verification}
To validate the above concept, we set up a near-field focusing array under the same conditions as those in Fig.~3 and performed electromagnetic simulations using the FEKO 2023. The near-field solver in FEKO was employed to obtain the electric fields corresponding to different focal scanning positions. The electric field distributions along the focal plane for the three different array types are shown in Fig.~6. Each dipole element has a length of 25~mm and a diameter of 0.1~mm, while the patch antenna has a length of 15.3~mm and a width of 23~mm, with a feed point offset by 7~mm from the center along the length direction. The substrate is a dielectric material with a thickness of 1.5~mm and a relative permittivity of 2.2. The center-to-center spacing of all elements is set to $2.27\lambda$.
From the simulation results, it is observed that for the vertically polarized dipole array, the beamwidth increases progressively during scanning, while the sidelobe level decreases. In contrast, the horizontally polarized dipole array and the planar patch array exhibit nearly identical scanning characteristics (with slight differences in peak gain), where both the main lobe beamwidth and the first sidelobe level remain nearly constant.
In all three cases, grating lobes appear at the edges of the scanning strip, confirming that this spacing value serves as a critical threshold for achieving complete scanning coverage. In practical applications, the element spacing can be slightly reduced below this threshold to prevent grating lobes from entering the scanning strip and to mitigate focal point deviations.

\begin{figure}[t!]
    \centering
    \includegraphics[width=0.78\linewidth]{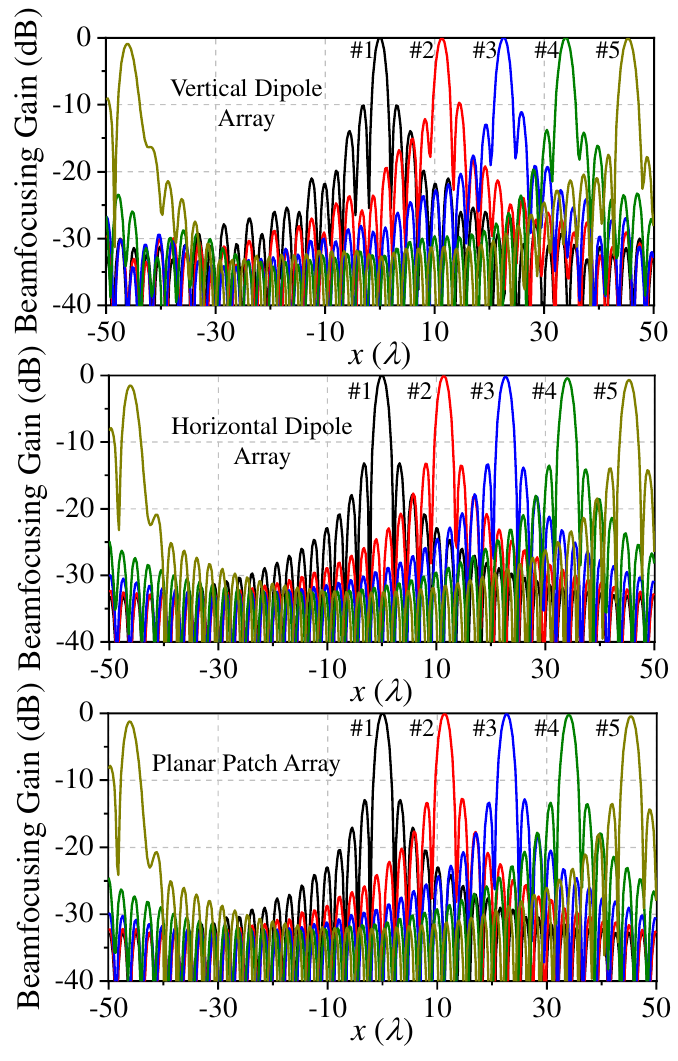} % Replace 'figure.png' with your image file
    \caption{Beam cross-sections along the focal direction for three types of antennas
.}\label{fig:example}
\end{figure}

\section{Conclusions}
In near-field beam focusing with finite-sized arrays, focal shift significantly degrades focusing accuracy and energy efficiency. To address this challenge, we revisit the problem from a communication degrees of freedom perspective. We propose a novel sparse array design with optimized element spacing to effectively suppress focal shift and ensure stable beam control across scanning angles. Simulation results based on dipole and patch antennas validate the proposed method, showing uniform energy distribution and highly accurate focal positioning. Our findings provide new insights for the practical deployment of near-field systems.

\newpage

\vfill

\end{document}